\documentclass[twocolumn]{aastex631}
\newcommand{\mytilde}{\raise.19ex\hbox{$\scriptstyle\sim$}}

\usepackage{amsmath}
\usepackage{xcolor}
\usepackage[utf8]{inputenc}
\defcitealias{2021ApJ...914...73Z}{Z21}

\shorttitle{Redistributed radio jet in Abell 514}
\shortauthors{Lee et al.}

\graphicspath{{./}{figures/}}

\begin{document}

\title{Discovery of a large-scale bent radio jet in the merging cluster Abell 514}

\author{Wonki Lee}
\affiliation{Yonsei University, Department of Astronomy, Seoul, Republic of Korea}
\affiliation{Harvard-Smithsonian Center for Astrophysics, 60 Garden St., Cambridge, MA 02138, USA}

\author{John ZuHone}
\affiliation{Harvard-Smithsonian Center for Astrophysics, 60 Garden St., Cambridge, MA 02138, USA}

\author{M. James Jee}
\affiliation{Yonsei University, Department of Astronomy, Seoul, Republic of Korea}
\affiliation{Department of Physics, University of California, Davis, One Shields Avenue, Davis, CA 95616, USA}
\correspondingauthor{M. James Jee}
\email{wonki.lee@yonsei.ac.kr, mkjee@yonsei.ac.kr}

\author{Kim HyeongHan}
\affiliation{Yonsei University, Department of Astronomy, Seoul, Republic of Korea}

\author{Ruta Kale}
\affiliation{National Centre for Radio Astrophysics–Tata Institute of Fundamental Research, Ganeshkhind, Pune, Maharashtra, INDIA}

\author{Eunmo Ahn}
\affiliation{Yonsei University, Department of Astronomy, Seoul, Republic of Korea}

\begin{abstract}

We report a discovery of a large-scale bent radio jet in the merging galaxy cluster Abell 514 ($z = 0.071$). 
The radio emission originates from the two radio lobes of the AGN located near the center of the southern subcluster and extends towards the southern outskirts with multiple bends. Its peculiar morphology is characterized by a $400$-kpc ``bridge'', a $300$-kpc ``arc'', and a $400$-kpc ``tail'', which together contribute to its largest linear size of $\mytilde0.7$~Mpc. We find that both the flux and spectral features of the emission change with the distance from the AGN. Also, the ``bridge'' presents a $60\%$ polarized radio emission, which coincided with an X-ray cold front. Based on our multi-wavelength observations, we propose that A514 presents a clear case for the redistribution of an old AGN plasma due to merger-driven gas motions.
We support our interpretation with idealized cluster merger simulations employing a passive tracer field to represent cosmic-ray electrons and find that merger-driven motions can efficiently create a cloud of these particles in the cluster outskirts, which later can be re-accelerated by the cluster merger shock and produce radio relics.

\end{abstract}


\section{Introduction} \label{sec:intro}

Many radio jets in merging galaxy clusters present a characteristic bent radio morphology \citep[e.g.,][]{1994ApJ...423...94B}.
This bent morphology is believed to be the result of interactions with the ambient intracluster medium 
\citep[ICM,][]{1976ApJ...205L...1O,1998Sci...280..400B}, 
which can bend \citep[e.g.,][]{2011ApJ...730...22P,2019ApJ...876..154N,2019ApJ...885...80N} or stretch the radio jets \citep[e.g.,][hereafter \citetalias{2021ApJ...914...73Z}]{2021Natur.593...47C,2021ApJ...914...73Z}. 
As the jets diffuse into the ICM, the non-thermal plasma loses its energy via synchrotron radiation and inverse-Compton scattering \citep[][]{1962SvA.....6..317K}. 
Although these energy losses remove the high-energy ($\gamma\sim10^4$, where $\gamma$ is the Lorentz factor) cosmic ray electrons (CRe), the jets are still rich in low-energy ($\gamma\lesssim10^3$) CRe.
This population of old, low-energy plasma, called fossil CRe,
has been hypothesized as the source of diffuse cluster radio emission through various energization processes \citep{2019SSRv..215...16V}. 
For example, if a cluster merger shock passes the cloud of fossil CRe that resides in the cluster outskirts, the diffusive shock acceleration creates a radio relic, an elongated arc-shaped radio emission feature found in the cluster periphery \citep[e.g.,][]{2016ApJ...823...13K}.
 
Recently, the unprecedented sensitivity at low-frequencies enabled by LOFAR \citep{2013A&A...556A...2V}, uGMRT \citep{2017CSci..113..707G} and MWA \citep{2013PASA...30....7T}, started to directly probe the fossil CRe content injected by the radio jets \citep[e.g.,][]{2017SciA....3E1634D,2021ApJ...909..198H,2021NatAs...5.1261B}. 
Nevertheless, the long-term spatial evolution of these jets and their contribution to the diffuse cluster radio emissions still remain unclear 
\citep[e.g.,][]{2023Galax..11...45V}. 

\begin{figure*}
    \centering
    \includegraphics[width=1.7\columnwidth]{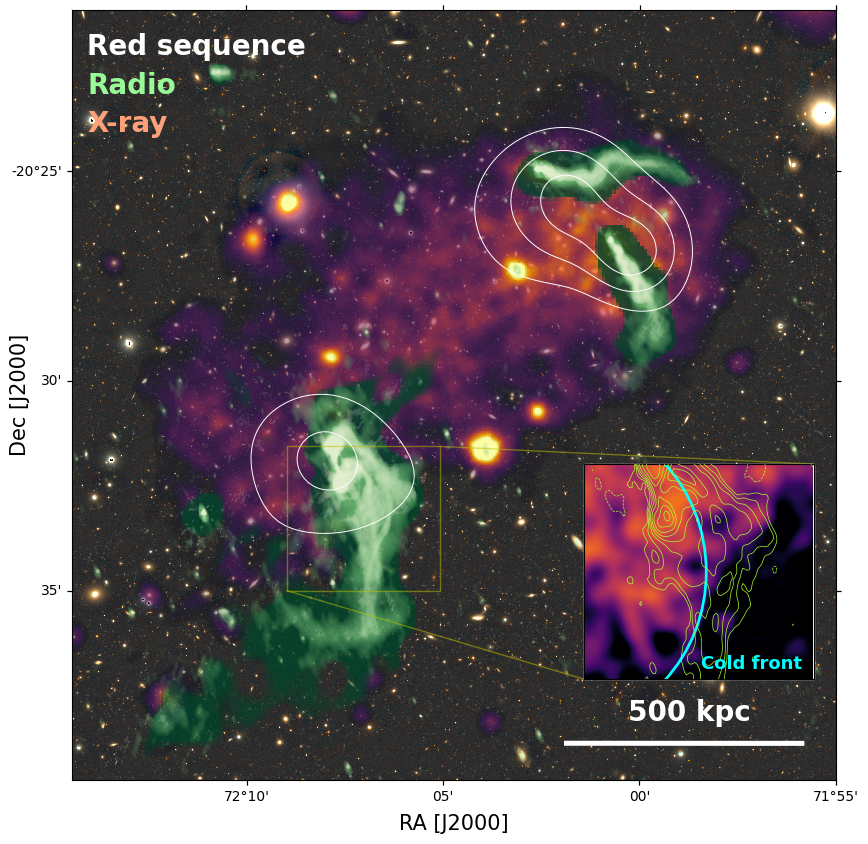}
    \caption{Multi-wavelength view of Abell 514.
    uGMRT radio maps (green), adaptively smoothed XMM-\textit{Newton} X-ray (red), and the red sequence number density contours are overlaid on the Magellan/Megacam optical color composite image.
    The radio map at a lower frequency is described with a darker green color scale. 
    The number density contours mark the level of 1.7, 2.3, and $3.0 \rm~galaxy~arcmin^{-2}$.
    The elongated X-ray and the two red sequence number density peaks suggest the collision between the northwest and the southeast clusters, while the southern AGN shows a large-scale bent radio jet with the largest linear size $\mytilde0.7\rm~Mpc$.
    The inset figure shows the point-source subtracted X-ray map around the radio features of the southern AGN. The green $410\rm~MHz$ radio contours start from $5\sigma$ and are spaced by a factor of 2. The cyan arc marks the X-ray surface brightness discontinuity, which we interpret as a cold front (\textsection\ref{sec:merger-driven}). 
    }
    \label{fig:title}
\end{figure*}
 
In this letter, we report a discovery of a large-scale bent radio jet in the merging cluster Abell 514 (hereafter A514). 
A514 is a massive ($M_{500
}\sim3\times10^{14}M_{\odot}$\footnote{Derived with $L_{X,\,[0.1,2.4]\rm~keV}=1.44\times10^{44}~\rm erg\,s^{-1}$
\citep{1996MNRAS.281..799E} and the X-ray luminosity-mass relation of \citet{2009A&A...498..361P}.}), low-redshift cluster merger at $z=0.0714$  \citep{1996ApJ...473..670F}.
\citet{1994ApJ...423...94B} reported the presence of three radio tails inside extended X-ray emission.
\citet{2008A&A...490..537W} claimed the detection of an X-ray discontinuity and suggested an ongoing collision in the NW-SE direction. 
In this study, we performed deep low-frequency radio observations using uGMRT and discovered a large-scale bent radio jet of A514. 
Based on our multi-wavelength data and simulations, we suggest that the observed radio feature represents old AGN plasma that has been redistributed by the merger-driven gas motions and evolved into a cloud of fossil CRe located in the outskirts of the cluster.

This letter is organized as follows. 
We outline the reduction process of the data in \textsection\ref{sec:observations} and present the radio features and their multi-wavelength properties in \textsection\ref{sec:general}.
We describe the numerical simulations of A514 in \textsection\ref{sec:discussion} before we summarize our findings in \textsection\ref{sec:conclusions}.  
We adopt a $\Lambda$CDM cosmology with $H_0=70$ km~s$^{-1}$Mpc$^{-1}$, $\Omega_m=0.3$, and $\Omega_{\Lambda}=0.7$. The angular size of $1\arcmin$ corresponds to a length scale of $\mytilde 80$~kpc at the cluster redshift $z=0.0714$. 

\section{Data \& Reduction} \label{sec:observations}
\subsection{Radio: uGMRT} \label{sec:radio}
We observed A514 using uGMRT Band 2 ($125-250 \rm \,MHz$), Band 3 ($250-500 \rm \,MHz$), and Band 4 ($550-850 \rm \,MHz$) with on-source integration times 
of $3.5$, $3.7$, and $4.3$ hours, respectively\footnote{Program code: $\rm 38\_124$, $\rm 41\_076$}.
We used 3C147 as the amplitude calibrator in all observations and selected 0447-220 (Band 2), 0521-207 (Band 3), and 0409-179 (Band 4) for the phase calibration.
We processed our uGMRT data with {\tt CASA v6.5.3}\footnote{http://casa.nrao.edu} and the latest {\tt CAPTURE} pipeline \citep{2021ExA....51...95K}. 
To briefly summarize, we self-calibrated the image using {\tt tclean}, implementing the multi-scale multi-frequency synthesis \citep[{\tt MS-MFS},][]{Rau2011} and {\tt W-Projection} \citep{Cornwell2008}. 
We employed the {\tt auto-multithresh} algorithm \citep{Kepley2020} to define the mask regions and the {\tt wbpbgmrt} module \footnote{https://github.com/ruta-k/uGMRTprimarybeam} to correct the primary beam.
We refer readers to \citet{2021ExA....51...95K} for the detailed reduction steps. 

\begin{figure*}
    \centering
    \includegraphics[width=2.\columnwidth]{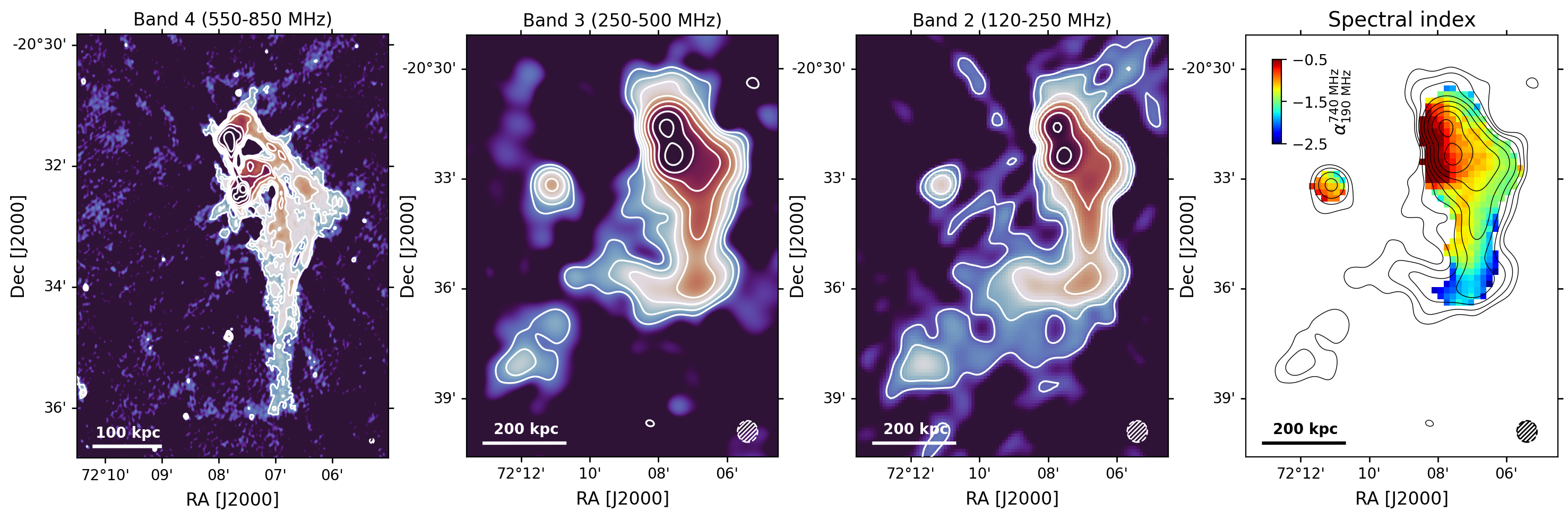}
    \caption{Radio maps of PKS 0446-20 at $740\rm~MHz$ (Band 4, $550-850\rm~MHz$, left) and smoothed radio maps with the beam size of $35''$ at $410\rm~MHz$ (Band 3, $250-500\rm~MHz$, middle left) and $190\rm~MHz$ (Band 2, $120-250\rm~MHz$, middle right). The right panel depicts the spectral index map derived from the three frequency channels. The radio maps are smoothed to match the beam size of $35''$. The annotated ellipse in the bottom right corner describes the synthesized beam size of each radio map. The overlaid contours mark the $5\sigma$ detection in the $740\rm~MHz$ and $410\rm~MHz$ radio maps and the $3\sigma$ detection in the $190\rm~MHz$ radio map. The contours are spaced by a factor of 2. 
    The radio emission gets fainter and presents a steeper spectrum with increasing distance from the AGN.}
    \label{fig:A514S}
\end{figure*}

We used the {\tt Briggs} weighting scheme to create the radio map. 
After testing different robust parameters, we set the robust parameter to 0 for Band 3 and 4 and 2 for Band2 in order to resolve both the substructures and the diffuse radio emission effectively. 
The noise level of the resulting radio map is $\sigma_{190\rm~MHz} \sim 3.0~\rm{mJy~b}^{-1}$ ($b_{190\rm~MHz}=33''\times21''$), $\sigma_{400\rm~MHz} \sim 0.08~\rm{mJy~b}^{-1}$ ($b_{400\rm~MHz}=12''\times5''$), and $\sigma_{740\rm~MHz} \sim 0.04~\rm{mJy~b}^{-1}$ ($b_{740\rm~MHz}=5''\times4''$) at Band 2, 3, and 4, respectively. 
We assume a flux calibration uncertainty of $10\%$ for Band 3 and 4 and $20\%$ for Band 2 \citep{2004ApJ...612..974C}.

\subsection{X-ray: XMM-{\it Newton}}
\label{sec:X-ray-calibration}
We retrieved the archival XMM-{\it Newton} observations of A514 (ObsId: 0142240101, 0142240201, and 0300920101),
and performed standard calibration based on the XMM-{\it Newton} Cookbook\footnote{http://heasarc.gsfc.nasa.gov/docs/xmm/esas/cookbook/} using the XMM-{\it Newton} Extended Source Analysis Software \citep[XMM-ESAS; SAS version 18.0.0;][]{2004ASPC..314..759G}.
After filtering and removing flaring events, the total clean exposure times of MOS1, MOS2, and PN are 14.8, 16.4, and 7.1 ks, respectively.
We extracted a soft-band (0.5-2 keV) image and masked the point sources detected by {\tt wavdetect} \citep{2002ApJS..138..185F} to obtain the surface brightness profile.
It was rebinned to have a minimum S/N $> 5$, using the {\tt PROFFIT v1.5} package \citep{2011A&A...526A..79E} for the analysis.

\subsection{Optical: Magellan/Megacam}\label{sec:opt}
We investigated the cluster galaxy distributions using optical data from
the Magellan/Megacam $g$ and $r$-filter observations (PI: W. Lee). The CCD level processing including overscan correction, trimming, and bad pixel removal was carried out at the Smithsonian Astrophysical Observatory (SAO) Telescope Data Center using the SAO Megacam reduction pipeline. We created sky flats using the same A514 observation consisting of multiple dithers and rotations. Astrometric corrections were obtained with the {\tt SCAMP} software \citep{2006ASPC..351..112B}. The final mosaic images were created with the {\tt SWARP} package \citep{2002ASPC..281..228B}. We ran {\tt SExtractor} \citep{1996A&AS..117..393B} in dual image mode using the $r$-band mosaic as the detection image to identify sources and obtain photometry. We calibrated the zero points using the PanSTARRS catalogs covering our field.

\section{A large-scale bent radio jet in Abell 514}
\label{sec:general}
Figure \ref{fig:title} presents the multi-wavelength view of A514.
The presence of the two galaxy number density peaks and the elongated X-ray morphology between them strongly suggest that A514 is currently undergoing a binary merger in the NW-SE direction. 
Previous observations have reported three head-tail radio galaxies in A514 \citep{1994ApJ...423...94B,1998MNRAS.301..609B,2001A&A...379..807G}, which are all classified as spectroscopic members \citep{1995AJ....109...14O,1998A&AS..129..399K}. 
With our new uGMRT radio observations, we revealed the bent morphology in these three radio galaxies. 
The two radio galaxies in the northern subcluster extend by $\mytilde300\rm~kpc$, before they bend at the position of the northern X-ray discontinuity reported by \citet{2008A&A...490..537W}.
Thus, they might present the case for jet bending caused by discontinuous ICM velocity across the front 
\citep[e.g., ][]{2011ApJ...730...22P}.

In this letter, we focus on the southern radio emission originating from the AGN PKS 0446-20 (hereafter A514S), which exhibits a remarkably peculiar morphology. 
\citet{2001A&A...379..807G} have reported two radio lobes with a hint of extended radio emission towards the south.
We found that this radio feature extends much longer than what has been reported.
The two radio lobes are connected to the $\mytilde400$-kpc long N-S structure that we refer to as ``\textit{bridge}''.
The southern end of the bridge meets the $\mytilde300$-kpc long ``\textit{arc}'', which is concave toward the cluster center. 
Finally, the eastern end of the arc touches the northern end of the $\mytilde400$-kpc long tail. 
The largest projected linear size of the combination of the three radio segments is $\mytilde0.7~$Mpc. 
Note that the aforementioned radio features, albeit in low resolution, can also be identified in the archival low-frequency radio survey data from TGSS \citep{2017A&A...598A..78I} and GLEAM \citep[e.g.,][]{2015PASA...32...25W}.

Figure \ref{fig:A514S} depicts the details of the radio emission in A514S at different frequencies.  
In the $740\rm~MHz$ radio map, the bridge is resolved into two filamentary substructures, whose surface brightness decreases rapidly toward the south.
On the other hand, the surface brightness level is elevated at the intersection of the bridge and the arc in the $410\rm~MHz$ and $190\rm~MHz$ radio maps.
The arc is only marginally detected in the $740\rm~MHz$ radio map while its presence is clear in the $410\rm~MHz$ and $190\rm~MHz$ radio maps.
The current data do not hint at any filamentary substructure within the arc.
Similarly, the tail presents a low surface brightness in the low-frequency radio maps.
The tail extends along the hypothesized merger axis of A514. 

\begin{table*}
	\centering
	\caption{Radio properties of the substructures in the southern bent jet}
	\label{obs}
	\begin{tabular}{cccccc} 
	    \hline
	    \hline
		 Name & $S_{740\rm~MHz}$ & $S_{400\rm~MHz}$ & $S_{190\rm~MHz}$ & $\alpha^{740\rm~MHz}_{190\rm~MHz}$ & $\alpha^{400\rm~MHz}_{190\rm~MHz}$\\
		    & $\rm Jy$ & $\rm Jy$  & $\rm Jy$ & & \\ 
		\hline
	    \textit{bridge} & $0.39\pm0.04$  & $0.84\pm0.08$  & $2.1\pm0.4$  & $-1.3\pm0.03$ & $-1.2\pm0.3$\\
		\textit{arc}    & $0.021\pm0.007$ & $0.10\pm0.01$  & $0.46\pm0.11$& $-2.6\pm0.1$ & $-1.9\pm0.3$\\
            \textit{tail}   & $-$    & $0.035\pm0.006$ & $0.26\pm0.08$  & $-$    & $-2.6\pm0.5$\\
        \hline
        \\
        \label{tab:flux}
	\end{tabular}

\end{table*}

\subsection{AGN Activity or Radio Relics?}
\label{sec:A514_spec}
Table \ref{tab:flux} summarizes the radio properties of the substructures in the southern bent jet. 
The integrated flux gradually decreases with distance from A514S in the order of the bridge, the arc, and the tail. 
We estimate the spectral slopes by fitting a power-law spectrum ($S\propto\nu^{\alpha}$) on the smoothed radio maps with the matched beam size of $35''$. The resulting spectral map in the right panel of Figure \ref{fig:A514S} clearly shows that the radio spectrum steepens with the distance from A514S.
Moreover, the spectrum of the arc is curved ($\alpha^{740\rm~MHz}_{190\rm~MHz}<\alpha^{400\rm~MHz}_{190\rm~MHz}$), whereas the bridge presents a flat spectrum ($\alpha^{740\rm~MHz}_{190\rm~MHz}\sim\alpha^{400\rm~MHz}_{190\rm~MHz}$). 
These patterns of changes in the flux and spectral slope are
anticipated due to the aging of the CRe originating from the radio lobes through synchrotron and inverse-Compton energy losses \citep[e.g.,][]{1962SvA.....6..317K,1973A&A....26..423J}. 

Based on its morphology, orientation, and location, one may identify the arc as a possible radio relic. 
Also, when we estimate its $1.4\rm~GHz$ luminosity assuming a power-law spectrum, we find that the value is in line with the expectation from the typical size-luminosity relation of radio relics \citep[][]{2012A&ARv..20...54F}. 
However, the steep and curved radio spectrum of the arc is different from the feature of typical radio relics, which are well-described with a single power-law description. 
A radio relic can feature a curved spectrum {\it if} a shock illuminates a cloud of fossil CRe that has been re-accelerated $>100\rm~Myrs$ ago \citep[e.g.,][]{2016ApJ...823...13K}.
Nevertheless, we believe that this scenario does not apply to the current case, where we expect the shock that re-accelerated the arc to propagate and re-accelerate the tail as well, because the tail extends along the collision axis.
As the spectral slope of the tail is steep, we conclude that the arc is not a relic-type source.

The aforementioned features of the AGN radio jet have been reported in other clusters.
A perpendicular reorientation of the radio jets similar to the bridge-arc intersection has been observed 
in the Coma cluster \citep{2020AJ....160..161L}, A3560 \citep{2013A&A...558A.146V}, A3376 \citep{2021Natur.593...47C}, and A3528 \citep{2018A&A...620A..25D}. 
As in the bridge of A514, filamentary substructures associated with the cluster AGN has been discovered in A4038 \citep{2018MNRAS.480.5352K}, A85 \citep{2022MNRAS.515.2245R}, A194 \citep{2022ApJ...935..168R}, and A3562 \citep{2022ApJ...934...49G}.
These examples demonstrate that the plasma ejected through AGN jets can undergo redistribution and evolve into peculiar radio morphologies, as observed in A514. 

\subsection{Jet Redistribution by merger-driven gas motions?}
\label{sec:merger-driven}

If the extended radio emission of A514 indeed originated from AGN activity, the AGN plasma must have been transported from the injection point at the AGN to the observed locations. 
One possible source of jet redistribution is the motion of the host galaxy A514S. 
If the host galaxy falls into the cluster with a high velocity, its jet will lag behind it. Also, the interaction with ICM can further warp the AGN jet distribution
\citep[e.g., ][]{2011ApJ...730...22P}. 
Another possible scenario is the redistribution of the jet material due to merger-driven gas motions (Z21). 
As described earlier, the rotational motion of the ICM can stretch the AGN bubble along a cold front given the fast motions underneath the front.

\begin{figure}
    \centering
    \includegraphics[width=\columnwidth]{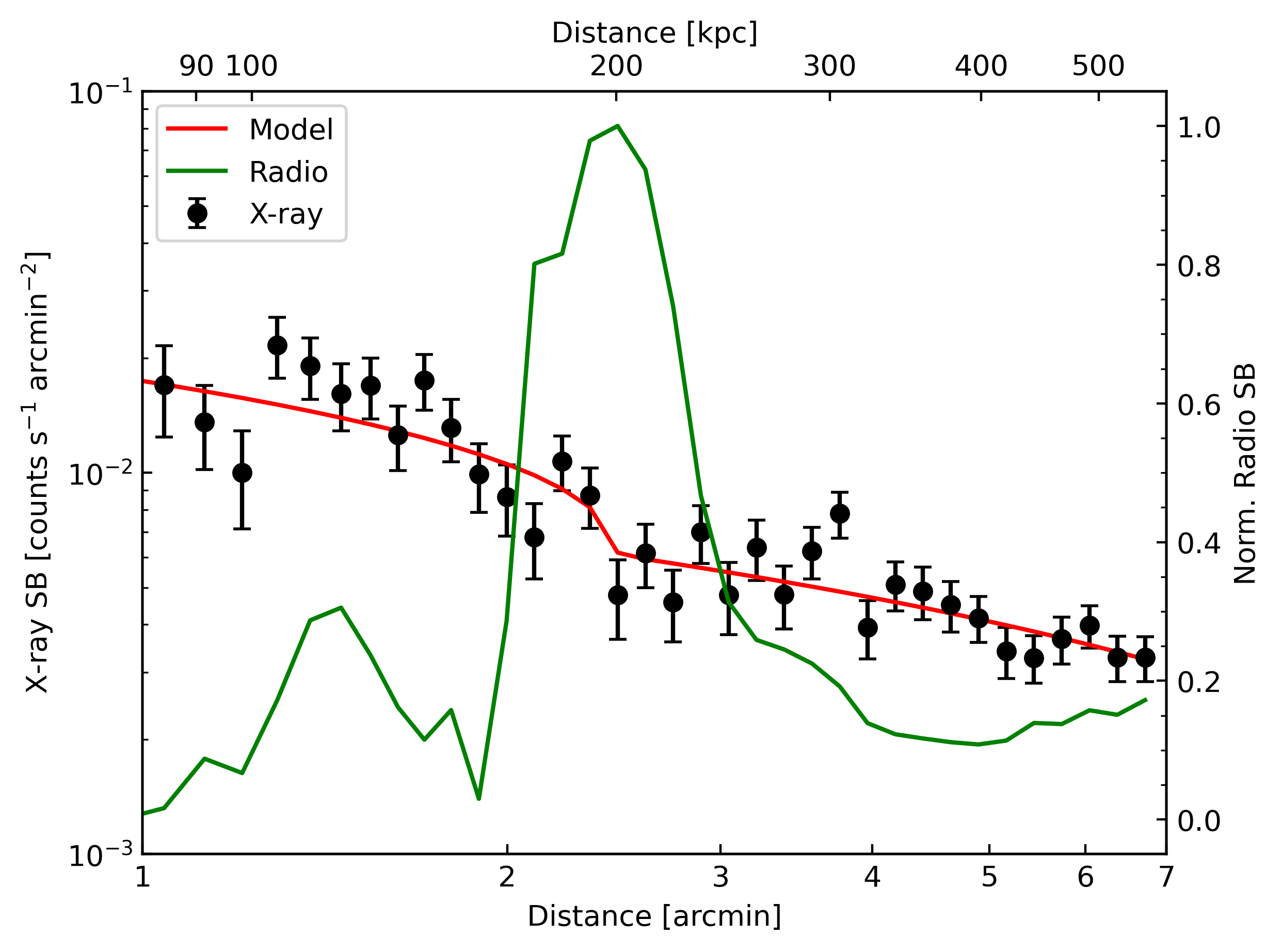}
    \caption{X-ray surface brightness profile (black) over the \textit{bridge}. 
    The X-ray surface brightness profile is well-described with a broken-power law (red) with a compression ratio $C\sim1.75\pm0.25$. 
    The radio surface brightness profile (green) presents a peak at the X-ray surface brightness jump. 
    }
    \label{fig:A514S-coldfront}
\end{figure}

Guided by the prediction of \citetalias{2021ApJ...914...73Z}, we assess the presence of a cold front by examining the X-ray surface brightness profile.
Figure \ref{fig:A514S-coldfront} presents an X-ray surface brightness discontinuity, where the surface brightness jump coincides with the radio emission from the bridge. 
We find that a broken power-law model ($\chi_{\rm red}^{2}\sim0.8$) is preferred over a single power-law model ($\chi_{\rm red}^{2}\sim1.6$). The fitted broken power-law model gives a compression ratio $C\sim1.75\pm0.25$.

This X-ray discontinuity can either be a cold front or a shock with $\mathcal{M}\sim1.5\pm0.2$ based on the Rankine-Hugoniot conditions based on the density jump alone. 
The current X-ray data cannot constrain the temperature of the western side (it corresponds to the upstream if the discontinuity is a shock) because of insufficient photon statistics. 
Still, since the discontinuity is found near the southern X-ray core, we believe that a cold front is more plausible than a merger shock that is launched ahead of the X-ray cores \citep[e.g.,][]{2000ApJ...541..542M}. 
Also, if the discontinuity originates from a shock, its shock propagation direction is in tension with the hypothesized merger axis deduced from the elongation of the ICM and galaxy distribution. 
The possibility of an equatorial shock is ruled out because it is expected to be found in the cluster periphery \citep[e.g.,][]{2018ApJ...857...26H}. 

\citetalias{2021ApJ...914...73Z} argued that the AGN jet redistributed by sloshing would provide a highly polarized radio emission due to the aligned and amplified intracluster magnetic fields along the cold front \citep[e.g.,][]{2011ApJ...743...16Z}. 
We examined if the \textit{bridge} is polarized with the NVSS data \citep{1998AJ....115.1693C}.
Interestingly, we find that the emission in the \textit{bridge} remains highly polarized ($\lesssim60\%$) despite the depolarization due to the large beam of NVSS ($45''$). 
Based on these signatures, we suggest that the \textit{bridge} traces the older material redistributed from the AGN jet along the cold front by the sloshing motion  \citepalias{2021ApJ...914...73Z}.

One can interpret the high polarization fraction and the filamentary substructures as originating from the magnetic flux bundles in the ICM, which may present extended radio features with the CRe diffusion along the magnetic fields \citep[e.g.,][]{2022ApJ...935..168R}. 
The direct comparison between the magnetic fields and the radio features is difficult with the current data, and we plan to address this in the follow-up project with MeerKAT L-band polarization observations. 

\section{Formation scenario of the bent radio jet} \label{sec:discussion}

We perform magnetohydrodynamics simulations of a binary cluster merger to test our scenario in explaining the observed radio morphology.
We set up the initial conditions of the simulations using the \texttt{cluster\_generator} package\footnote{https://github.com/jzuhone/cluster\_generator}. 
We place the two clusters inside a $15~$Mpc closed box, with each cluster composed of a dark halo modelled by an NFW profile\citep{Navarro1996} and the ICM modelled by a modified beta profile ICM \citep{2006ApJ...640..691V}.
We set a turbulent magnetic field and normalize its strength so that the ICM has an average ratio between the thermal and the magnetic pressure of $\mytilde100$. 
We guide the readers to \citet{2019ApJ...883..118B} for more details on the initial setups. 

\begin{figure}
    \centering
    \includegraphics[width=0.8\columnwidth]{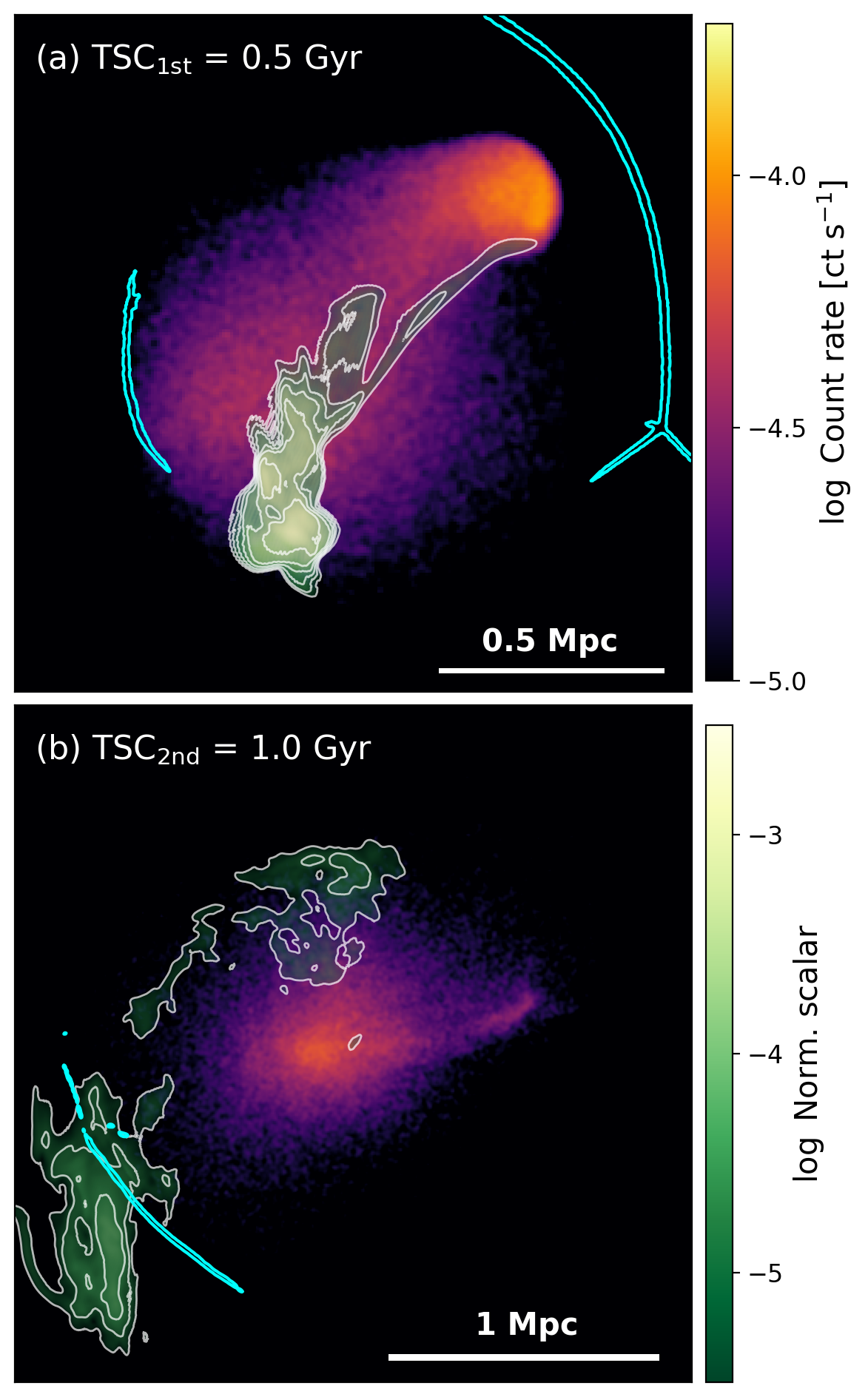}
    \caption{
    Mock X-ray surface brightness map of the simulations overlaid with scalar field density map (green), which is normalized with the injected density. 
    The scalar field contours (white) start at $10^{-5}$ and are spaced equally by a factor of 2.
    The cyan arcs mark the temperature discontinuity, which shows the shock fronts. 
    At $\mytilde0.5\rm~Gyr$ after the first closest passage (top panel), the bubble extends by $\mytilde0.4$ Mpc in a direction similar to Abell 514.
    At $\mytilde1.0\rm~Gyr$ after the second closest passage ($\mytilde4.5\rm~Gyr$ after the first closest passage, bottom panel), the bubble stretches along the merger shock in the cluster outskirts.
    }
    \label{fig:simulation}
\end{figure}

We perform the merger simulations with \texttt{GAMER-2} \citep[][]{2018MNRAS.481.4815S}, a GPU-accelerated Adaptive MEsh Refinement code.
The code refines the patch down to $\mytilde8~$kpc based on the number of dark matter particles. 
In a suite of simulations with different collision parameters, we create a mock X-ray map using the \texttt{yt}\citep{2011ApJS..192....9T} and \texttt{pyXSIM} packages \citep{2016ascl.soft08002Z} and determine the viewing angle that can reproduce the observed X-ray core separation of $\mytilde0.7\rm~Mpc$. 
We find that an off-axis collision between two clusters with masses of $2$ and $1\times10^{14}~M_{\odot}$ clusters can reproduce the overall X-ray features of A514 (top panel of Figure \ref{fig:simulation}). This collision started with a relative infall velocity of $900\rm~km~s^{-1}$, an initial separation of $3\rm~Mpc$, and an initial impact parameter of $600\rm~kpc$. The snapshot and viewing angle which best match the features of A514 are $\mytilde0.5$~Gyr after the first core passage and a line-of-sight angle of $\mytilde70\deg$\footnote{We define the viewing angle of plane of sky merger as $90\deg$.}.
We note that the northern X-ray core in A514 is the less-massive cluster in this scenario.

In this collision, we use a passive scalar to track gas sampled inside a randomly-placed $50$ kpc sphere before the first closest passage. 
We use this bubble gas to describe the redistribution of fossil CRe injected by a cluster AGN. 
This passive tracer is a simple model, which does not include a physical model that can modify the substructures of the bubble or the ICM itself \citep[e.g.,][]{2021Galax...9...91Z}. 
However, we assume the spatial extent of the redistributed bubble can roughly describe the extent of the redistributed non-thermal plasma as the non-thermal plasma is known to follow the ICM gas flows after it has become mixed with the thermal plasma \citep[e.g.,][]{2023arXiv230207881B}. 

Fiugre \ref{fig:simulation} demonstrates that the bubble redistributes along the cold front formed ahead of the infalling main cluster, where the cluster core stretches the bubble perpendicular to its motion.
This infalling motion stretches the bubble by $\mytilde400~$kpc in the north-south direction with a curved morphology, like the \textit{bridge} in A514, which supports our scenario for the jet redistribution by the merger-driven gas motions.  

\subsection{Injection of Fossil CRe in the Cluster Outskirts} \label{sec:future}

The diffuse radio emission of A514 extends towards the cluster periphery, which implies that the AGN plasma will distribute in the cluster outskirts and possibly seed radio relics.
To check the long-term evolution of AGN plasma, we further perform the simulation until the second closest passage. 

The bottom panel of Figure \ref{fig:simulation} presents the simulated merger $\mytilde1.0\rm~Gyr$ after the second closest passage (i.e., $\mytilde 4.5\rm~Gyr$ after the first closest passage).
At this late merger phase, the bubble plasma has reached the cluster outskirts, stretched by $\mytilde1$ Mpc, and extends tangentially with respect to the cluster center. 
These characteristics of the redistributed bubble plasma are not surprising as the cold front, which was the primary source of the jet redistribution, is known to create an arc-shape gas tail that overruns its parent cluster's dark matter during an off-axis cluster merger \citep[e.g.,][]{2019ApJ...874..112S,2020ApJ...894...60L}. 
As a result, we find that the merger shock from the second closest passage aligns with the redistributed bubble plasma (see the cyan arc in Figure \ref{fig:simulation}). Under this condition, we expect this bubble to be re-accelerated and present a radio relic-like emission.

The fossil CRe injection presented in Figure \ref{fig:simulation} is similar to the scenario described in  \citetalias{2021ApJ...914...73Z}. 
Our results suggest that the redistribution process can be efficient. 
Specifically, our simulation shows that the AGN plasma injected before the first closest passage can be transformed into a fossil CRe cloud before the merger shock from the second closest passage reaches the plasma.  

Accurate modeling of the CRe is required to predict the characteristics of the radio relics created under this scenario, which is beyond the scope of this work. 
Instead, we qualitatively discuss the observable signatures of the radio relics formed under this scenario. 
One possible signature is a steep radio halo. 
The turbulence generated from the cluster merger can re-accelerate the non-thermal plasma inside the ICM and present a radio halo \citep{2001MNRAS.320..365B}.
As the bubble plasma redistributes 1 Gyr after the second closest passage, we expect the radio halo would have a steep radio spectrum \citep[e.g.,][]{2013MNRAS.429.3564D}.
Another possible feature is the inhomogeneous surface brightness of the radio relic.
The rotational motion that redistributes the bubble plasma is tangential to the cluster center, aligning the bubble with the merger shock. 
However, the merger shock does not need to pass the center of the redistributed bubble plasma.
As a result, the re-accelerated radio relic will produce an inhomogeneous surface brightness feature with the brightest emission produced where the fossil CRe exist. 

\section{Summary} \label{sec:conclusions}

We have reported the discovery of a large-scale bent radio jet in the merging cluster A514. The radio emission originates from the two radio lobes of the AGN located in the central region of the southern cluster and extends towards the southern periphery. 
The entirety of the radio feature is believed to be traced back to the AGN because both its flux and spectral slope decrease as the distance from the AGN increases.
We have detected both a discontinuity in X-ray surface brightness and high polarization at the location of the extended radio emission. 
We interpret these features as a result of the jet plasma redistribution along the cold front of the recent cluster merger.

To assess the viability of our scenario, we conducted cluster merger simulations using a toy model. 
The simulations show that a passive plasma bubble, injected during an off-axis cluster merger, can undergo stretching along the cold front of the infalling cluster. This stretching process results in an extended radio emission that resembles the observation in A514.
At the late merger phase, the bubble redistributes at the cluster outskirt with its elongation aligned tangential to the cluster. 
Based on these findings, we propose that A514 shows the early stage of fossil CRe injection, where the redistributed fossil CRe in the cluster outskirts can help boost the acceleration efficiency of the merger shocks.
    
\hfill \break

M. J. Jee acknowledges support for the current research from the National Research Foundation (NRF) of Korea under the programs 2022R1A2C1003130 and RS-2023-00219959.
RK acknowledges the support of the Department of Atomic Energy, Government of India, under project no. 12-R\&D-TFR-5.02-0700.

\bibliography{reference}{}
\bibliographystyle{aasjournal}

\end{document}